\documentclass{article}
\usepackage{spconf,amsmath,graphicx,tabularx, booktabs, amssymb}
\usepackage{multirow}
\newcolumntype{C}[1]{>{\PreserveBackslash\centering}p{#1}}
\newcolumntype{R}[1]{>{\PreserveBackslash\raggedleft}p{#1}}
\newcolumntype{L}[1]{>{\PreserveBackslash\raggedright}p{#1}}

\title{SPGM: Prioritizing Local Features for enhanced speech separation performance}
\name{\begin{tabular}[t]{c} Jia Qi Yip$^1$$^2$, Shengkui Zhao$^1$, Yukun Ma$^1$, Chongjia Ni$^1$, Chong Zhang$^1$, Hao Wang$^1$, \\ \textit{Trung Hieu Nguyen$^1$, Kun Zhou$^1$, Dianwen Ng$^1$$^2$, Eng Siong Chng$^2$, Bin Ma$^1$} \end{tabular}}
\address{
  $^1$Alibaba Group\\
  $^2$Nanyang Technological University (NTU), Singapore\\
  }

\begin{document}
%
\maketitle
\vspace{-15pt}
\begin{abstract}
Dual-path is a popular architecture for speech separation models (e.g. Sepformer) which splits long sequences into overlapping chunks for its intra- and inter-blocks that separately model intra-chunk local features and inter-chunk global relationships. However, it has been found that inter-blocks, which comprise half a dual-path model's parameters, contribute minimally to performance. Thus, we propose the Single-Path Global Modulation (SPGM) block to replace inter-blocks. SPGM is named after its structure consisting of a parameter-free global pooling module followed by a modulation module comprising only 2\% of the model's total parameters. The SPGM block allows all transformer layers in the model to be dedicated to local feature modelling, making the overall model single-path. SPGM achieves 22.1 dB SI-SDRi on WSJ0-2Mix and 20.4 dB SI-SDRi on Libri2Mix, exceeding the performance of Sepformer by 0.5 dB and 0.3 dB respectively and matches the performance of recent SOTA models with up to 8 times fewer parameters. Model and weights are available at huggingface.co/yipjiaqi/spgm
\end{abstract}
\vspace{-10pt}
\section{Introduction}
\label{sec:intro}
\vspace{-5pt}

Single-channel Speech Separation (SS) is the task of obtaining clean, single-speaker speech from a speech mixture of multiple overlapping speakers. To manage the long sequence lengths necessitated by the waveform-to-waveform prediction task of SS, models reduce the sequence using UNet-like or dual-path architectures. UNet-like architectures championed by ConvTasNet~\cite{convtasnet} consist of a series of layers that model the feature sequence at different time scales before recombining them for the final output. The dual-path architecture championed by DPRNN~\cite{dprnn} splits the long input into a series of overlapping chunks and uses an intra-block to model local features within a chunk, followed by an inter-block of the same size to model global features across chunks. Both approaches have remained popular, with UNet-like models~\cite{tzinis2020sudo}~\cite{Hu2021afrcnn}~\cite{li2023tdanet}~\cite{chen2023s4m} generally delivering better efficiency while dual-path models~\cite{2023sepformer}~\cite{Lutati2022SepItAA}~\cite{Lutati2023SeparateAD} continue to break performance records.

\begin{figure}[t]
  \centering
  \includegraphics[width=0.90\linewidth]{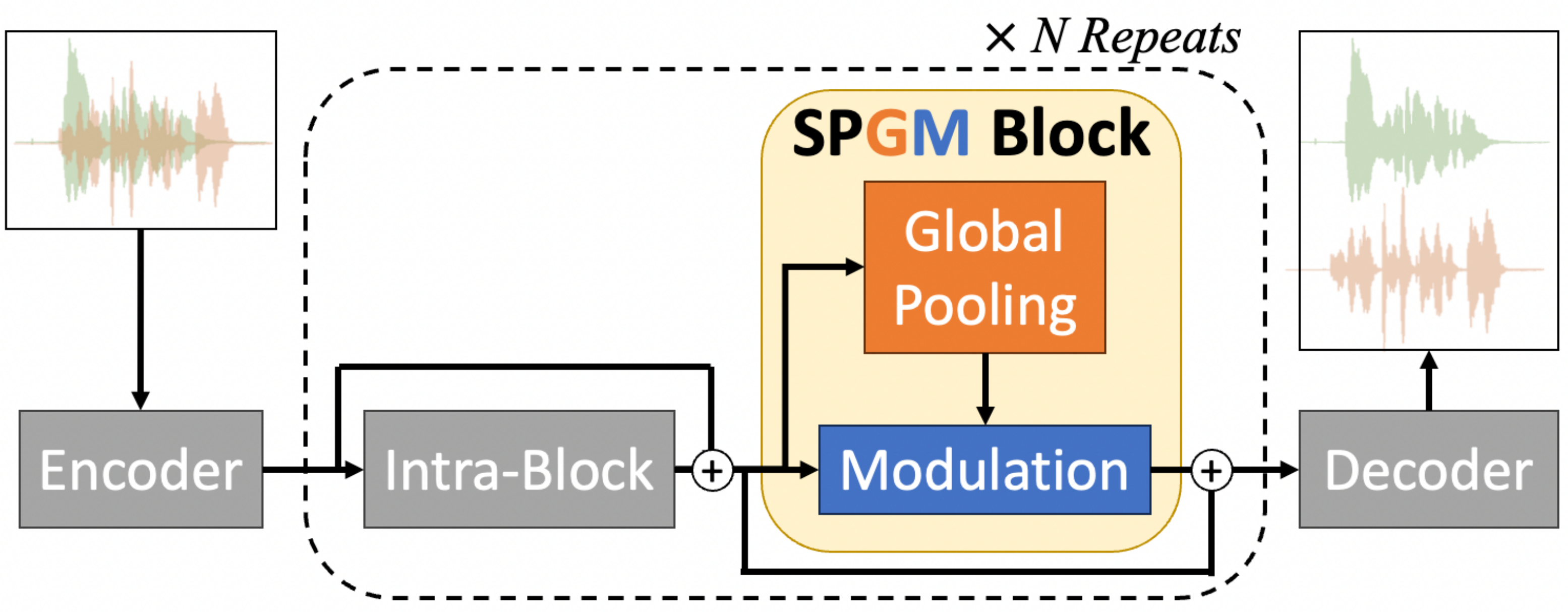}
  \caption{Overview of the proposed SPGM model. The SPGM block (yellow box) replaces an inter-block and represents our key contribution.}
  \label{fig:SPGM-overview}
  \vspace{-15pt}
\end{figure}

Given the impressive results achieved by dual-path models, some models have also sought to improve the dual-path architecture~\cite{papez}~\cite{rixen2022qdpn}~\cite{zhao2023mossformer}. A known issue with the dual-path architecture is that the inter-block comprises half of the model's total parameters but does not deliver commensurate performance~\cite{2023sepformer}. Recently, \cite{papez} validated this by training an auxiliary model to adaptively prune tokens from Sepformer that did not contribute to performance. While there was minimal pruning in the intra-block, most of the inter-block features were pruned. This suggests that the detailed modeling by transformers in the inter-block is excessive. 

Thus, \cite{papez} replaced the inter-block with chunk-averaged memory tokens between each intra-block, creating an efficient model at the cost of poorer performance. Meanwhile, QDPN~\cite{rixen2022qdpn} used deep down- and up-sampling layers in the inter-block to efficiently extract global features, delivering good performance but with significantly increased model size.

In this paper, we address efficient global modeling in dual-path SS models through the Single-Path Global Modulation (SPGM) block. SPGM performs global pooling followed by feature modulation, as shown in Figure~\ref{fig:SPGM-overview}, requiring only 0.5M trainable parameters in 2 linear layers that scale with embedding size. SPGM's simple and efficient global modeling enables additional intra-block layers, improving performance. SPGM achieves an SI-SDRi of 22.1 dB on the WSJ0-2Mix dataset and 20.4 dB on the Libri2Mix dataset, outperforming the strong baseline of Sepformer by 0.5 dB and 0.3 dB respectively. SPGM also matches the performance of SOTA models, QDPN~\cite{rixen2022qdpn} and SFSRNet~\cite{sfsrnet}, using fewer parameters.

\vspace{-5pt}
\section{Methodology} 
\vspace{-5pt}
\subsection{Model Architecture}
\vspace{-5pt}

\begin{figure}[t]
  \centering
  \includegraphics[width=0.5\linewidth]{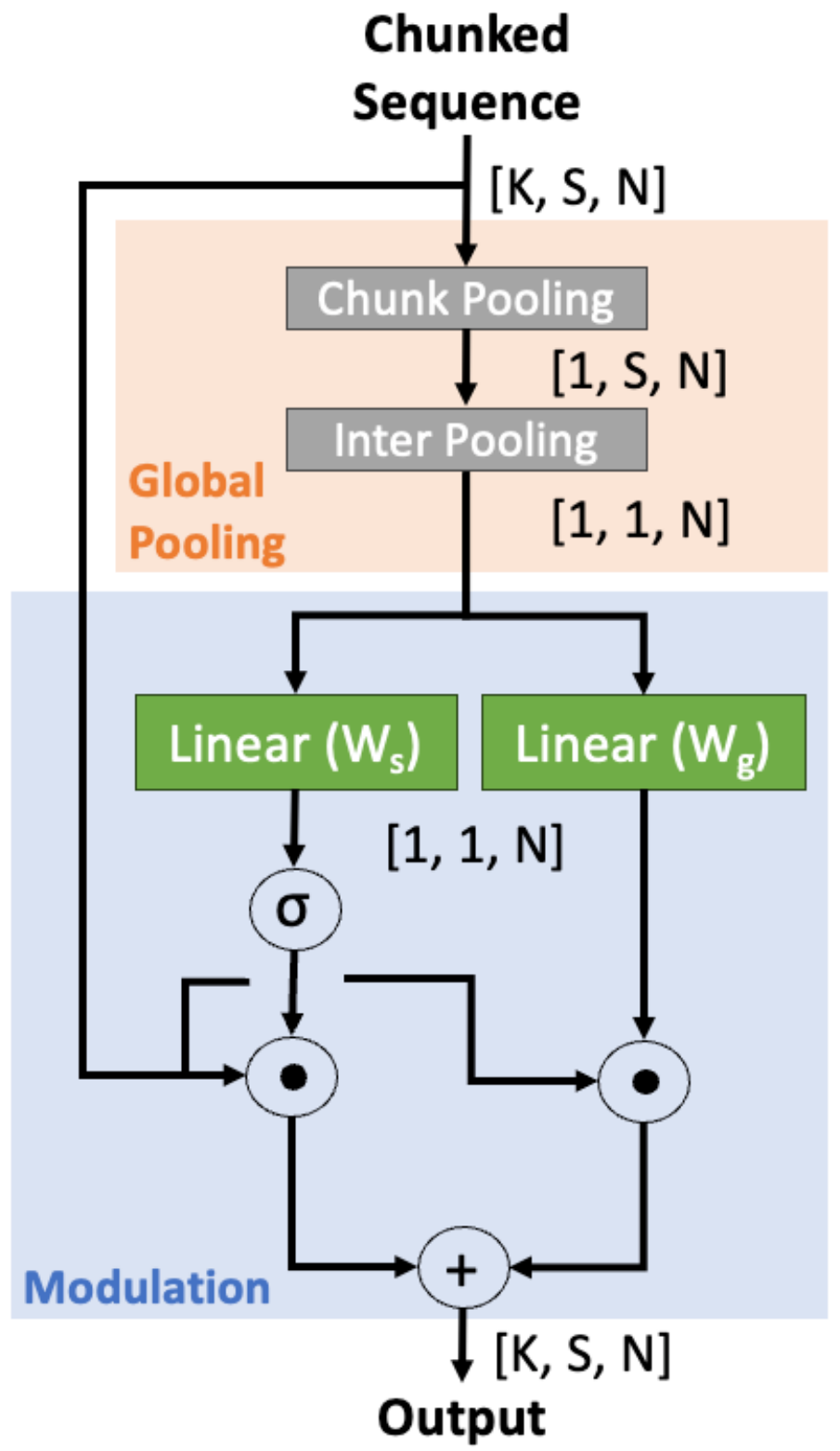}
  \caption{The SPGM block consists of the global pooling module (orange) and the modulation module (blue). K is the number of time steps, S is the number of chunks and N is the embedding size. Refer to Equation~\ref{eqn:modulationEqn} for the detailed implementation of the modulation module.}
  \label{fig:poolmod}
  \vspace{-10pt}
\end{figure}

The design of SPGM, shown in Figure~\ref{fig:SPGM-overview}, aims to model global information as efficiently as possible so that more model layers can be dedicated to local feature modeling. This is driven by work~\cite{2023sepformer}~\cite{papez} which showed how the Sepformer inter-blocks contribute minimally to overall performance, from which we infer that the global information required by the model across chunks is relatively simple and does not need to be modeled using transformer layers. 

To model global information, a block would need to perform 2 tasks. Firstly, the model must be able to model global, time-independent information from across the feature sequence. Secondly, it must pass the global information back to the entire feature sequence. In SPGM, detailed in Figure~\ref{fig:poolmod}, the former is achieved in the pooling module through a parameter-free pooling operation while the latter is achieved in the modulation module through a mechanism which requires only 2 trainable linear layers. 

The rest of the model as shown in Figure~\ref{fig:SPGM-overview} consists of standard SS model components. The Encoder and Decoder are a single layer of 1D convolution and 1D transposed convolution respectively. Refer to Section~\ref{sec:model_configuration} for implementation details of the model.

\begin{figure}[t]
  \centering
  \includegraphics[width=\linewidth]{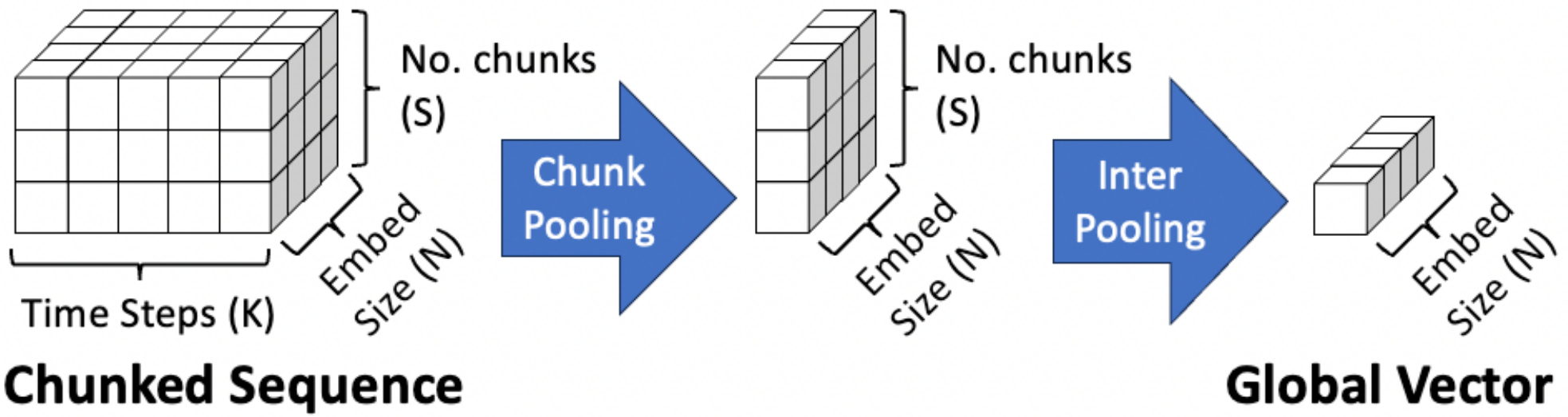}
  \caption{Illustration of the change in dimensions across the chunk and inter pooling process in the global pooling module.}
  \label{fig:PoolingProcess}
  \vspace{-10pt}
\end{figure}

\vspace{-5pt}
\subsection{Single Path Global Modulation}
\vspace{-5pt}

The SPGM pooling module, shown in the orange box in Figure~\ref{fig:poolmod}, consists of chunk pooling and inter pooling. The overall change in dimensions across the SPGM pooling module is further illustrated in Figure~\ref{fig:PoolingProcess}. Chunk pooling obtains a single vector of embedding size from each chunk. Inter pooling is a per-channel average and is taken across the chunk dimension, resulting in a single global vector of embedding size. This global vector is then used as the input for the SPGM modulation module. We experiment with different chunk pooling methods, which we discuss in Section~\ref{sec:poolmethod}, while inter pooling is always done using simple averaging to give each chunk an equal weight in the final global vector.

The SPGM modulation module, shown in the blue box in Figure~\ref{fig:poolmod}, accepts the features from the intra-block and the global vector. The modulation aims to condition each feature in the sequence by a time-independent global vector so that subsequent intra-blocks can refine the local features based on the imbued global information. The modulation block also has to do this with minimal computation and parameters. 

The SPGM modulation module is designed based on feature-wise linear modulation~\cite{perez2018film} which has been used to fuse multi-scale information in recent SS models like TDANet~\cite{li2023tdanet} and S4M~\cite{chen2023s4m} to good effect and is thus highly suitable for its role in SPGM as well.

During the forward pass, the modulation layer refines the global vector from the pooling module. Our method makes use of a linear and non-linear path, each differentiated with a simple $\mathcal{O}(N)$ linear layer in each path, where N is the channel size of the model. Modulation is achieved through element-wise multiplication of the embedding vectors and the feature sequence.

The modulation operation can be expressed as the following equation:
\begin{equation}
\label{eqn:modulationEqn}
\begin{aligned}
    x_{o} &= \sigma (W_{s}x_{emb}^{T}) \otimes x_{f} + W_{g}x_{emb}^{T}\otimes x_{f}
\end{aligned}
\end{equation} 

\noindent where $x_{emb} \in \mathbb{R}^{N}$ is the [1,1,N] global vector from the global pooling module, $W_{s}$ and $W_{g}$ are the weights of linear layers, as shown in Figure~\ref{fig:poolmod}, each with size $N \times N$, and $x_{f}, x_{o} \in \mathbb{R}^{K\times S\times N}$ are the feature input and output of the SPGM block respectively. $\sigma$ represents the sigmoid activation function. $K$ is the number of time steps, $S$ is the number of chunks and $N$ is the embedding size.

\begin{figure}[t]
  \centering
  \includegraphics[width=0.5\linewidth]{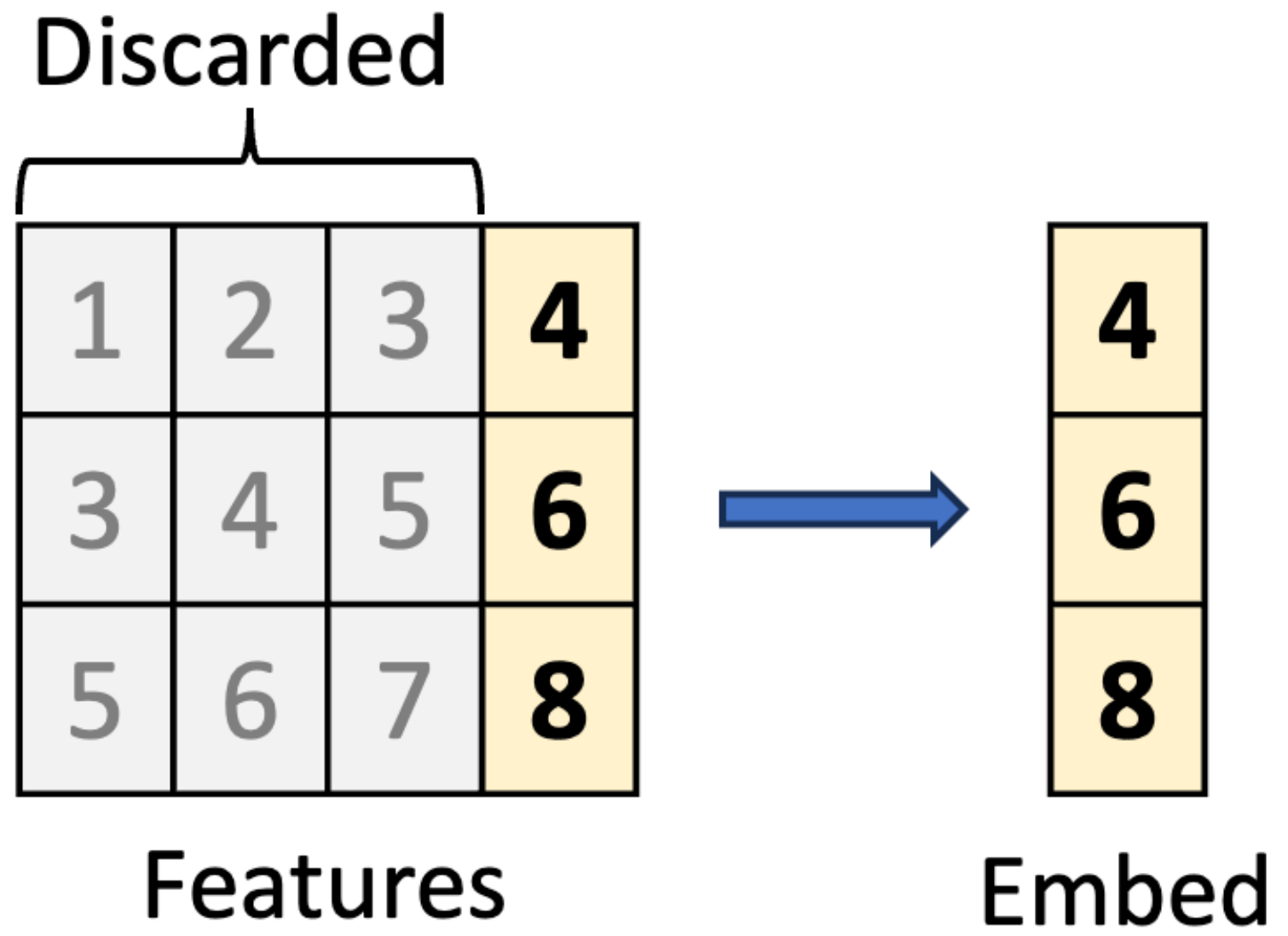}
  \caption{Illustration of the last element selection (LE) pooling method using a chunk size of 4 with a 50\% overlap on a single channel. LE selects the last element of each chunk as the global vector while the remaining features are not used to derive the global embedding.}
  \label{fig:PoolingMethods}
  \vspace{-10pt}
\end{figure}

\vspace{-5pt}
\subsection{Pooling Methods}
\vspace{-5pt}
\label{sec:poolmethod}
In this work we experiment with two different pooling methods for chunk pooling in the SPGM global pooling module: last element selection and attentive pooling.

Last Element Selection (LE) refers to the selection of the last element of each chunk as the global vector as shown in Figure~\ref{fig:PoolingMethods}. This works because the model creates overlapping chunks, so the last element of each chunk would be redundant and can be repurposed by the preceding transformer layer as a specialized global vector. Through the attention mechanism, the intra-block transformer layers can adaptively store global information in the last element, which is both initialized by and superimposed on the feature sequence. Furthermore, this method requires no additional trainable parameters. Unlike the recently proposed Papez~\cite{papez}, which utilizes memory tokens in a single-path model, our method does not require an explicit memory token, eliminating the need for additional hyperparameter tuning over the random initialization of additional memory parameters. 

Attentive Pooling (AP) seeks to develop an adaptive weight for each feature in a chunk for aggregation. It passes the features through a linear layer which outputs a single value for the weight of the feature in the final aggregated vector. Our implementation follows the attentive pooling method commonly used in speaker verification models~\cite{safari2020attnPool}. Although this method is not strictly parameter-free, the parameters required for this pooling method are exactly 256 since it only requires a single linear layer to map an input of embedding size (256) to a single value, which is negligible in the context of a 26M parameter model. 

\vspace{-5pt}
\section{Experiments}
\vspace{-5pt}
\subsection{Datasets}
\vspace{-5pt}

The WSJ0-2Mix~\cite{wsj0mix} and Libri2Mix~\cite{cosentino2020librimix} datasets are standard benchmarks for SS. The WSJ0-2Mix dataset consists of mixtures drawn from the WSJ0 corpus. The Libri2Mix dataset is generated from the LibriSpeech train-clean-360 Corpus. In our experiments, both datasets have a sampling rate of 8kHz. Additionally, we use 10s cropping on the WSJ0-2Mix dataset and 5s cropping on the Libri2Mix dataset.

\subsection{Model Configuration}
\label{sec:model_configuration}
The original Sepformer's dual-path architecture consisted of intra- and inter- blocks consisting of 8 transformer layers each. Each Sepformer block consists of 1 intra-block and 1 inter-block. The model has 2 Sepformer blocks in total: 16 intra-transformer layers and 16 inter-transformer layers.

To enable direct comparisons, we train as baseline 2 intra-only Sepformer variants, labeled IntraSepformer, with 16 and 32 intra-transformer layers respectively. Our models, SPGM-x-S and SPGM-x (x denoting the chunk pooling method of either LE or AP), also have 16 and 32 Intra-transformer layers respectively, producing the 4 models reported in Table~\ref{tab:size-ablation}.

The encoder and decoder of all the models each have a kernel size of 16 and stride of 8. In all transformer layers, the number of channels is 256, while the size of the hidden feed-forward network is 1024. The number of heads in each transformer layer is 8.

\vspace{-5pt}
\subsection{Training Parameters}
All model training is performed using the Speechbrain framework~\cite{speechbrain}. We trained for a maximum of 200 epochs with a starting learning rate of $1.5e^{-4}$ using the Adam optimizer. The learning rate is halved with patience of 3. Training is stopped when the maximum number of epochs is reached or when the learning rate reaches a minimum of $1.0e^{-8}$. During training, data augmentation using Speed Perturbation of a random value between 95\% and 105\% is used. The utterance-level Permutation Invariant Training loss~\cite{uPIT} was used to update the weights of the model during training and the performance of the model is measured using the standard Scale-invariant signal-to-distortion ratio improvement (SI-SDRi)~\cite{USS}.

\vspace{-5pt}
\section{Results}
\vspace{-5pt}
\subsection{Effectiveness of the SPGM block}
\vspace{-5pt}

The results of the experiments reported in in Table~\ref{tab:size-ablation} demonstrate that although local feature modeling is important, some global information modeling is required, which can be provided using the SPGM block. Comparing the SPGM-x-S models, we see that the addition of SPGM blocks significantly improves performance over the IntraSepformer model variants by up to 2.1 dB on WSJ0-2mix and 1.4 dB on Libri2Mix. The SPGM-x model variants also outperform their IntraSepformer counterparts. This shows that naively increasing the number of Intra transformer layers without global modeling does not lead to better performance.

Among the different chunk pooling methods used on the SPGM-x-S models, Attentive Pooling performs best on the Libri2Mix dataset, while the Last Element Selection performs best on the WSJ02Mix dataset. However, in the case of the WSJ0-2Mix dataset, the performance gap of 0.1dB is smaller than the performance gap of 0.4dB seen on the Libri2Mix dataset. 

When the number of intra-transformer layers is increased, the effect of the chunk pooling mechanism disappears and we observe that SPGM-LE and SPGM-Att have identical performance of 22.1 dB on WSJ02Mix and 20.4 dB on Libri2Mix. This could suggest that the attention pooling was performed implicitly within the Intra-layers itself and did not need to be done in the SPGM block. The models achieve the same performance regardless of pooling method because with the increased depth of the model, the increased number of layers allows the last element to receive global information from the chunk on par with the attention mechanism used in attentive pooling.

\vspace{-10pt}
\begin{table}[h!]
  \caption{Comparison of Sepformer variants with 16 intra-layers and 32 intra-layers against SPGM models with the same number of intra-layers and SPGM with different chunk pooling mechanisms. The datasets used are WSJ02Mix (W2J) and Libri2Mix (L2M). LE stands for Last Element selection and AP stands for Attention pooling used in the pooling module of the SPGM block used.}
  \vspace{5pt}
  \label{tab:size-ablation}
  \centering
    \begin{tabular}{ |p{2cm}|c|c|c|c|c|  }
     \hline
    \multicolumn{1}{|c|}{\multirow{2}{*}{Model}}
    & \multicolumn{1}{c|}{\multirow{2}{*}{\begin{tabular}[c]{@{}c@{}}Param \\ (M)\end{tabular}}} 
    & \multicolumn{1}{c|}{\multirow{2}{*}{Intra}}
    & \multicolumn{1}{c|}{\multirow{2}{*}{Inter}}
    & \multicolumn{2}{c|}{SI-SDRi (dB)} \\ \cline{5-6} 
    
    \multicolumn{1}{|c|}{} & \multicolumn{1}{c|}{} & \multicolumn{1}{c|}{}& \multicolumn{1}{c|}{}& \multicolumn{1}{c|}{WJ2} 
    & \multicolumn{1}{c|}{L2M} \\
    \hline
    Sepformer~\cite{2023sepformer}          & 25.7  & 16 & 16 & 21.6 & 20.1  \\
    \hline
    IntraSepformer          & 13.0  & 16 & 0  & 18.7 & 18.2\\
    SPGM-LE-S          & 13.3  & 16 & LE & 20.8 & 19.2 \\
    SPGM-AP-S          & 13.3  & 16 & AP & 20.7 & 19.6\\
    \hline
    IntraSepformer          & 25.7  & 32 & 0  & 19.6 & 18.8  \\
    SPGM-LE            & 26.2  & 32 & LE & 22.1 & 20.4 \\
    SPGM-AP            & 26.2  & 32 & AP & 22.1 & 20.4 \\
    \hline
    \end{tabular}
    \vspace{-10pt}
\end{table}

\vspace{-10pt}
\subsection{Comparison with Recent Models}
The results of the SPGM model in comparison with recent state-of-the-art models are reported in Table~\ref{tab:comparisons}. Since SPGM-LE and SPGM-AP have the same performance, they are not differentiated here. On the Libri2Mix dataset, our model achieves the same performance of 20.4dB SI-SDRi as the much larger SFSRNet~\cite{sfsrnet}. On the WSJ0-2Mix dataset, our model achieves the performance of 22.1dB SI-SDRi, on par with QDPN~\cite{rixen2022qdpn}, which is also a very large model with 8 times the number of parameters. It is also worth noting that TF-GridNet~\cite{tfgrid} achieves 23.5dB. However, it is a time-frequency domain model operating on additional information not available to the time-domain models in Table~\ref{tab:comparisons}.

Compared with the original Sepformer model, we achieved a performance improvement of 0.5dB on WSJ0-2Mix and 0.3dB on Libri2Mix with only a 0.5M increase in the number of parameters\footnote{These 0.5M parameters are the parameters of the 4 SPGM blocks which each have two sets of linear weights resulting in $4 \times 2 \times 256 \times 256 = 524,288$ parameters. This number of parameters is a function of the embedding size used in the encoder and decoder, which in this case is 256.}. This is a modest increase in the number of parameters given the performance improvement, especially in comparison with SFSRNet and QDPN, which achieve the same performance improvement relative to Sepformer, but have at least double the parameters. 

Additionally, we compute the MACs (Multiply-Accumulate Operations) using PyTorch-OpCounter\footnote{https://github.com/Lyken17/pytorch-OpCounter} for the various models trained in this work to assess the additional computation created by the SPGM block. The SPGM-block has a negligible impact on the overall model which is dominated by the MACs arising from the transformer layers. For example, on a 1-second sequence, each SPGM-block contributes 0.02 GMACs while the overall model requires 77 GMACs.

\begin{table}[t!]
\vspace{-5pt}
  \caption{Performance of SPGM models on the WSJ0-2Mix and Libri2Mix datasets in comparison with Sepformer and other systems.}
  \vspace{5pt}
  \label{tab:comparisons}
  \centering
    \begin{tabular}{ |p{2.8cm}|c|c|c|  }
     \hline
    \multicolumn{1}{|c|}{\multirow{2}{*}{Model}} & \multicolumn{1}{c|}{\multirow{2}{*}{\begin{tabular}[c]{@{}c@{}}Params \\ (M)\end{tabular}}} & \multicolumn{2}{c|}{SI-SDRi (dB)} \\ \cline{3-4} 
    \multicolumn{1}{|c|}{} & \multicolumn{1}{c|}{} & \multicolumn{1}{c|}{WSJ02Mix} & \multicolumn{1}{c|}{Libri2Mix} \\ \hline
    TDANet~\cite{li2023tdanet}          & 2.3     & 10.8 & - \\
    ConvTasNet~\cite{convtasnet}        & 5.1   & 15.3 & -\\
    SuDoRMRF~\cite{tzinis2020sudo}      & 2.7   & 17.0 & -\\
    DPRNN~\cite{dprnn}                  & 2.6   & 18.8 & -\\
    Papez~\cite{rixen2022qdpn}          & 1.47  & 19.2 & 17.2\\
    DPTNet~\cite{dptnet}                & 2.7   & 20.2 & 16.2\\
    Wavesplit~\cite{wavesplit}~\cite{2023sepformer} & 29 & 21.0 & 19.5 \\ 
    SFSRNet~\cite{sfsrnet}              & 59.0  & 22.0 & \textbf{20.4}\\ 
    QDPN~\cite{rixen2022qdpn}           & 200   & ~\textbf{22.1} & -\\
    \hline
    Sepformer~\cite{2023sepformer}      & 25.7 & 21.6 & 20.1\\
    SPGM (Ours)                      & 26.2  & \textbf{22.1} & \textbf{20.4}\\
    \hline
    \end{tabular}
\end{table}

\vspace{-5pt}
\section{Conclusion}
\vspace{-5pt}
We have shown here that local feature modeling is more important to the speech separation task, such that using a simple and efficient model for global modeling and reallocating parameters to local feature modeling is sufficient to surpass the performance of Sepformer. This allows SPGM to match the performance of large models while using significantly fewer parameters. We achieved SI-SDRi of 22.1 dB on WSJ02Mix and 20.4 dB on Libri2Mix with 26.2M parameters, matching the performance of models with 59M and 200M parameters and exceeding the performance of the original Sepformer. Additionally, global modulation is an obvious handle for auxilliary information in speaker extraction using efficient speaker embedding models~\cite{yip2023acanet}. Thus, we hope that SPGM can serve as a foundation for future work.

\vspace{-5pt}
\section{Acknowledgements}
\vspace{-5pt}
This work was supported by Alibaba Group through Alibaba Innovative Research (AIR) Program and Alibaba-NTU Singapore Joint Research Institute (JRI), Nanyang Technological University, Singapore. The computational work for this article was partially performed on resources of the National Supercomputing Centre, Singapore (https://www.nscc.sg).

\bibliographystyle{IEEEbib}
{\footnotesize
\bibliography{main}}

\end{document}